\documentclass[11pt,a4paper]{article}
\pdfoutput=1
\usepackage{jheppub}
\usepackage{amsmath,amscd}
\usepackage{slashed}
\usepackage{amsfonts}
\usepackage{amssymb}
\usepackage{graphicx}
\usepackage{color}
\usepackage[normalem]{ulem}
\usepackage{hyperref}

\newcommand{\RR}{\mathbb{R}} 
\newcommand{\ZZ}{\mathbb{Z}} 
\newcommand{\NN}{\mathbb{N}} 

\def\tr         {{\rm  tr}}
\def\cala         {{\cal A}}

\def\calc         {{\cal C}}
\def\cald         {{\cal D}}

\def\call         {{\cal L}}

\def\caln         {{\cal N}}

\newsavebox{\uuunit}
\sbox{\uuunit}
    {\setlength{\unitlength}{0.825em}
     \begin{picture}(0.6,0.7)
        \thinlines
        \put(0,0){\line(1,0){0.5}}
        \put(0.15,0){\line(0,1){0.7}}
        \put(0.35,0){\line(0,1){0.8}}
       \multiput(0.3,0.8)(-0.04,-0.02){12}{\rule{0.5pt}{0.5pt}}
     \end {picture}}

\def\be{\begin{equation}}
\def\ee{\end{equation}}
\def\bea{\begin{eqnarray}}
\def\eea{\end{eqnarray}}

\def\a{\alpha}
\def\b{\beta}
\def\h{\eta}
\def\g{\gamma}

\def\d{\delta}
\def\e{\epsilon}
\def\D{\Delta}
\def\l{\lambda}
\def\L{\Lambda}

\def\f{\phi}

\def\m{\mu}
\def\n{\nu}
\def\o{\omega}

\def\r{\rho}

\def\t{\tau}

\def\sF{{{ F}\!\!\!\!\hskip.8pt\hbox{\raise1pt\hbox{/}}\,}}
\def\som{{{ \omega}\!\!\!\!\hskip.8pt\hbox{\raise1pt\hbox{/}}\,}}
\def\sJ{{{\rm J}\!\!\!\!\hskip.8pt\hbox{\raise1pt\hbox{/}}\,}}

\def\F{\Phi}
\def\pa{\partial}

\def\to{\rightarrow}
\def\nonu{\nonumber \\{}}
\def\half{{1 \over 2}}

\title{Simple Unfolded Equations for Massive Higher Spins in AdS$_3$}
\author[a]{Pan Kessel}
\author[b]{and Joris Raeymaekers}

\affiliation[a]{Machine Learning Group, Technische Universit\"at, Berlin Marchstrasse 23, 10587 Berlin, Germany.}
\affiliation[b]{CEICO, Institute of Physics of the Czech Academy of Sciences,  Na Slovance 2, 182 21 Prague 8, Czech Republic.}
\emailAdd{joris@fzu.cz}
\emailAdd{pan.kessel@gmail.com}

\abstract{We propose a simple unfolded description of free  massive higher spin particles in anti-de-Sitter spacetime. 
 While	our unfolded equation of motion has the standard form of a covariant constancy condition, 
 our formulation differs from the standard one in that our field takes values in a different internal space, which for us is simply
  a unitary irreducible representation of the symmetry group.
 Our main result is the explicit construction,
	  for the case of AdS$_3$, of a map from our formulation to the standard  wave equations for massive higher spin particles, as well as to  the  unfolded description prevalent in the literature.
It is hoped that our formulation may be used to clarify the group-theoretic content of interactions in higher spin theories.}
\arxivnumber{}
\begin{document}
\maketitle

\section{Introduction and Summary}\label{secintro}
The study of relativistic wave equations, whose solution spaces carry unitary irreducible representations of the spacetime symmetry group,
lay at the birth of quantum field theory and was undertaken by several towering figures in the field.  For example, in 1939 Fierz and Pauli \cite{Fierz:1939ix} wrote  down equations describing free, massive particles transforming as a symmetric rank $s$ tensor:
\be
(\Box -M^2) \f_{\m_1 \ldots \m_s} =0, \qquad
\nabla^{\m}  \f_{\m \m_2 \ldots \m_s} =0 \, ,\label{FP}
\ee
where $\f_{\m_1 \ldots \m_s}$ is a totally symmetric traceless tensor. In the anti-de Sitter space AdS$_{d+1}$ the set of solutions of these equations, upon imposing suitable boundary conditions, form a unitary, irreducible   representation  $D(\D, s)$  of the symmetry algebra $so(2,d)$ with lowest energy $\D$, where
\be
M^2 = \D (\D  -d) - s \, .\label{FPmass}
\ee
 We refer to \cite{Rahman:2015pzl} for a modern review of  relativistic wave equations and references to the original literature.

	In recent years a different yet equivalent formulation of relativistic wave equations has proven  useful, namely  the unfolded formulation due to Vasiliev and collaborators (see \cite{Bekaert:2005vh} and \cite{Didenko:2014dwa} for reviews). In this formulation, covariant wave equations like (\ref{FP}) are replaced by a system of coupled first-order equations typically
containing an infinite number of auxiliary fields. A  beautiful feature of  unfolded equations  is  that they geometrize covariant wave equations like (\ref{FP}), since  they can be interpreted  as a covariant constancy condition on a  section of a certain vector bundle over  AdS$_{d+1}$.
Unfolded equations were first proposed  for massless higher spins in (A)dS$_{4}$ \cite{Fradkin:1987ks}, and subsequently generalized to other dimensions, massive fields  and flat backgrounds \cite{Lopatin:1987hz}-\cite{Buchbinder:2016jgk}. One advantage of the unfolded formulation is that it formally
facilitates the coupling of  massive  fields to massless higher spin gauge fields, and therefore it lies at the core of Vasiliev's construction of interacting higher spin theories \cite{Vasiliev:1990en},\cite{Prokushkin:1998bq}.
This has in turn played an important role in recently uncovered examples  of holographic duality, where bulk Vasiliev theories
were argued to be dual to holographic boundary CFTs possessing conserved currents of spin greater then two.

One example in the context of AdS$_3$/CFT$_2$ holography is the minimal model holography proposed by Gaberdiel and Gopakumar  \cite{Gaberdiel:2010pz}.
Here, the bulk theory contains a massive scalar field with mass $m^2 = \l^2 -1$, which is coupled to  massless higher spin fields with  $hs[\l ]$ gauge symmetry through  unfolded equations \cite{Vasiliev:1992gr},\cite{Barabanshchikov:1996mc}. A second example, which inspired the current work, is provided by the tensionless  limit of string theory on the $AdS_3 \times S^3 \times T_4$ background with Ramond-Ramond flux. In this case, the bulk theory contains massless higher spin fields with a gauge symmetry which goes under the name of the 'higher spin square' (HSS)\footnote{Recently, it was shown \cite{Giribet:2018ada},\cite{Gaberdiel:2018rqv} that the symmetric  orbifold CFT also describes a subsector of the tensionless limit  of string theory on the S-dual background with NS-NS flux.} \cite{Gaberdiel:2014cha},\cite{Gaberdiel:2015mra},\cite{Gaberdiel:2015wpo}.  Furthermore, the symmetric orbifold CFT contains many  spinning primaries with spins $s = |h - \bar h| \in \NN / 2$ which in the bulk correspond to massive higher spin fields. As in the example above, it is therefore desirable
to have an unfolded description of  the  massive higher spin equations (\ref{FP}). In \cite{Raeymaekers:2016mmm}, one of us proposed a linearized unfolded equation   describing the massive higher spin fields in the untwisted sector in the background of a HSS gauge field. When restricted to a pure spin-two background, one notices that this equation gives a different and, we feel, simpler unfolded description of massive HS fields than the ones in the literature.

In this work, we clarify the  relation between this unfolded formulation of massive higher spins in AdS$_3$  and the standard one. 
In both formulations,   the basic field $C(x)$ is a zero-form section of a vector bundle over AdS$_{3}$, taking values in an infinite-dimensional representation $R$ of the symmetry algebra $so(2,d)$. The field equations simply state that $C(x)$ is covariantly constant:
  \be
\left( d + \cala_R \right) C =0\, .\label{unfeqintr}
\ee
In this equation, $\cala$ stands for
 the flat AdS$_{d+1}$ connection made out of the vielbein and spin connection
\be
\cala = e^a P_a + \half \o^{ab} M_{ab}.\label{calaintr}
\ee
The subscript $R$ in (\ref{unfeqintr}) means that the generators are taken in the representation $R$. The equation (\ref{unfeqintr}) states that the general solution is obtained by picking an arbitrary vector  $C_0$ at the origin and parallel transporting it.  In
terms of the group element $G$ in writing  $ \cala = G^{-1} dG$,
the general solution is $C(x) = G_R^{-1} (x)  C_0$.

In the standard unfolded formulation \cite{Vasiliev:1992gr},\cite{Barabanshchikov:1996mc},\cite{Boulanger:2014vya} for a massive spin-$s$ field on AdS$_3$, the representation $R$ acts on basis vectors  $V^{(t)}_a$, with  $|a|\leq t$ and $t = s , s+1, \ldots$,  which for fixed $t$ transform as a spin-$t$ representation under the Lorentz subalgebra $so(1,2)$.  The  AdS translation generators act on these  as in formula (\ref{Pmaction}) below. In the  most widely known example describing a spin-0 field, the $V^{(t)}_a$ are generators of the higher spin algebra $hs[\l ]$ and the action of the AdS translation generators comes from the `lone-star' product. The resulting system of equations are equivalent to the Fierz-Pauli description (\ref{FP}), since one can prove a `central on mass-shell' theorem which states that the lowest spin-$s$ component of
$C$ is precisely the Fierz-Pauli field $\f_{\m_1 \ldots \m_s}$ above.

In this work, we will explore a different unfolded formulation, where the representation $R$ in (\ref{unfeqintr}) is instead simply taken to be the
unitary irreducible representation $D(\D , s)$ itself. One advantage of this formulation is that the space of solutions to (\ref{unfeqintr})  forms a Hilbert space with inner product
inherited from  $D(\D , s)$.  While for this choice for $R$ the fact that  (\ref{unfeqintr}) gives a field theory realization of $D(\D , s)$ is almost tautological, it is not a priori clear  if there is an analogue of the central on mass-shell  theorem allowing one to reconstruct the Fierz-Pauli field from $C$, nor how this unfolded formulation is related to the standard one. The main goal of this work is to address these questions. A key property is  that the generators $V^{(t)}_a$ of the standard unfolded formulation can be constructed as non-normalizeable state in the $D(\D , s)$ Hilbert space\footnote{In \cite{Iazeolla:2008ix}, a similar construction was performed for the case of  massless representations in AdS$_D$ with $D \ge 4$.}.
This  allows us to construct a linear map or `intertwiner' between the two representations, and  construct from our field the unfolded field of the standard formulation. The restriction to the spin-$s$ component of the latter then  leads to the desired on mass-shell theorem. As a corollary, our results allow for a completely algebraic construction of the mode solutions of the Fierz-Pauli equations, see equation
(\ref{modesol}) below.

 Since in the present unfolded  formulation, the group theoretic meaning is completely transparent and involves only the representation $D(\D , s)$,  it may be hoped that
 it may shed light on the group-theoretic content of the interaction vertices in Vasiliev theory. This may be of use in constructing as yet unknown interactions
 in the theory based on the higher spin square. As a first step towards such a construction,  we will show  in a separate publication how the equation proposed in \cite{Raeymaekers:2016mmm} combines an infinite set of our massive higher spin equations of the form (\ref{unfeqintr}) into a single multiplet of the higher spin square.

\section{Simple Unfolded Equations on AdS$_{d+1}$}

In this section, we review some aspects of the geometry  of AdS$_{d+1}$ and propose and analyze our unfolded equations.

\subsection{Coset Description}
We start out by recalling 
the coset description of  anti-de Sitter space.
The $d+1$-dimensional  anti-de Sitter spacetime AdS$_{d+1}$ can (up to global issues which are not relevant at present) be described as the homogenous symmetric space $SO(2, d)/SO(1,d)$, where the isotropy subgroup $SO(1,d)$ is the Lorentz group in $d+1$ dimensions. We will denote by $M_{AB},\ A,B = 0',0, \ldots d$ the generators of
the Lie algebra $so(2,d)$ with commutation relations
\be
[M_{AB},M_{CD}]=i \left( \h_{BC} M_{AD} - \h_{AC} M_{BD} - \h_{BD} M_{AC}+ \h_{AD} M_{BC}\right)\, ,\label{adsDalg}
\ee
where $\h_{AB} = {\rm diag } (--+ \ldots +)$.
 In a  unitary representation, the $M_{AB}$ are represented by Hermitian operators. We will split them up in `AdS translations', i.e. the coset generators $P_a = M_{0'a}, \ a = 0, \ldots d$, and Lorentz
 generators $M_{ab} \in SO(1,d)$. They satisfy\footnote{Note that we set the AdS radius to one.}
\bea
\,[P_a,P_b] &=& i M_{ab}\nonu
\,[M_{ab},P_c] &=& - 2 i \h_{c[a} P_{b]}\nonu
\,[M_{ab},M_{cd}]&=& - 2 i \h_{c[a} M_{b]d} + 2 i \h_{d[a} M_{b]d}\, .\label{AdSalgP}
\eea

 Due to  homogeneity, points in  AdS$_{d+1}$ can be viewed as coset representatives $G(x) \in SO(2, d)$, for example we could use a `canonical' parametrization where $G(x) = \exp x^a P_a $. The  symmetry group $SO(2, d)$ acts on the coset element as
 \be
 g G(x) = G(x') h, \qquad g \in SO(2,d), \qquad h\in SO(1,d).\label{groupaction}
 \ee
 The infinitesimal version of this relation, setting $g = 1 + \e^{AB} M_{AB} $, defines the Killing vectors $l_{AB}^\m \pa_\m$ through $x'^\m= x^\m - \e^{AB} l_{AB}^\m$ and
 what we will call the `Lorentz-compensator' fields $W^{ab}_{AB} $ through $h = 1 - \half \e^{AB} W_{AB}^{ab} M_{ab}$.  It follows from (\ref{groupaction}) that these satisfy
 \be
 G^{-1} M_{AB}G = - l_{AB}^\m G(x)^{-1} \pa_\m G(x) - \half W_{AB}^{ab} M_{ab}.\label{Liedef}
 \ee
 It can be shown that the Killing vectors $l^\m_{AB}$ obey  the same commutation relations (\ref{adsDalg}) as the generators $M_{AB}$. We refer to \cite{Castellani:1991et} for a  proof and a review of   the differential geometry of coset spaces.

From the coset representative we construct the flat $so(2,d)$-valued connection
\be
\cala = G^{-1}d G\, .\label{puregauge}
\ee
 It can be decomposed into vielbein and spin connection parts as follows
\be
\cala = e^a P_a + \half \o^{ab} M_{ab}\equiv e + \o.\label{vbsc}
\ee
Using this relation in (\ref{Liedef}), one finds an expression for the Killing vectors and Lorentz compensators in terms of the vielbein, spin connection and adjoint representation components of $G$:
\bea
l_{AB}^\m &=& - e_a^\m (G^{-1} M_{AB} G)^a\\
W_{AB}^{ab} &=&- (G^{-1} M_{AB} G)^{ab} -  l_{AB}^\m \o_\m^{ab}\label{KVcomp}.
\eea

\subsection{Unfolded equations}

Following Wigner's definition, a quantum mechanical particle can be identified with a unitary, irreducible representation of the spacetime symmetry group in which the energy is bounded from below.  In the case of AdS$_{d+1}$, particle representations are built on a set of primary states $|\D, {\bf s}\rangle$ which form a unitary  irreducible representation of the maximal compact subalgebra $so(2) \oplus so(d)$. Here, $\D$ is the eigenvalue of the energy operator $P_0$ while ${\bf s}$ denotes quantum numbers specifying a unitary irreducible representation of   $so(d)$. The states $|\D, {\bf s}\rangle$ are annihilated by the energy lowering operators
\be
J^-_a = M_{0a}  + i P_a \,.
\ee
The representation is  built up by acting on the states $|\D, {\bf s}\rangle$ with the energy raising operators $J^+_a = M_{0a}  - i P_a $ and will be denoted by $D(\D, {\bf s})$. If ${\bf s}$ is a totally symmetric rank-s tensor, the quadratic Casimir takes the value
\be
\half M_{AB}M^{AB} = \D( \D-d) + s(s+ d-2) .\label{symmCas}
\ee

By a field theory realization of the particle representation $D(\D, {\bf s})$, we mean a set of spacetime-dependent fields which  satisfy
a set of equations (and possibly boundary conditions) which are invariant under the spacetime isometry algebra, such that the solution space transforms as the representation $D(\D, {\bf s})$. For example, the Fierz-Pauli equations (\ref{FP}) in AdS$_{d+1}$ give, upon imposing suitable boundary conditions, a field theory realization of the representation\footnote{This statement holds only for $d>2$. For AdS$_3$, where the subgroup of spatial rotations reduces to $so(2)$,  we will review in Section \ref{sectopmass} below that the Fierz-Pauli equations describe two irreducible representations with opposite signs of the spatial $so(2)$ helicity.\label{fnads3}}   $D(\D, s)$, where $s$ stands for the symmetric rank
$s$ tensor representation.

As anticipated in the Section \ref{secintro}, we  will now show that an alternative field theory realization
of a  particle representation $R= D(\D, {\bf s})$ is provided by the system of  equations
\be
(d + \cala_R) C(x)  = 0,\label{adsD}
\ee
where $C(x)$ is a zero-form  which takes values in an internal space which is precisely the representation space $R$.  The connection $ \cala$ is the AdS$_{d+1}$ connection (\ref{vbsc}), and the subscript $R$ means that the generators in (\ref{vbsc}) are taken in the representation $R$. For notational simplicity, we will drop this subscript in what follows.
Note that the equation (\ref{adsD}) is integrable due to the fact that $\cala$ is a flat connection.

\subsection{Lorentz and Diffeomorphism Covariance}
Let us first show the covariance of the equations (\ref{adsD}) under diffeomorphisms and local Lorentz transformations.
For this, we observe that the equations are  gauge-invariant under local $SO(2,d)$ transformations under which both the background  $\cala$  and the field $C$ transform, in the following way:
\bea
\cala &\to &\L  (\cala + d )\L^{-1} \label{formalgauge1}\\
C &\to & \L C\label{formalgauge2}
\eea
When $\L = \exp (\l^{ab} (x) M_{ab})$ belongs to the Lorentz subgroup $SO(1,d)$, these transformations encode the covariance of the equation (\ref{adsD}) under local Lorentz transformations. Indeed, the first equation (\ref{formalgauge1}) implies, using the commutation relations (\ref{AdSalgP}), the standard transformation of the vielbein and spin connection under local Lorentz transformations.
The second equation (\ref{formalgauge2}) elucidates the Lorentz tranformation character of our master field $C$: it transforms as the representation $R$, decomposed under the Lorentz subalgebra
$so(1,d)$. Therefore, $C$ doesn't transform irreducibly under Lorentz tranformations in general, in contrast to e.g.  symmetric tensor field of the Fierz-Pauli system. We note that the equation  (\ref{adsD}) can be rewritten
as
\be
(\nabla + e^a P_a) C =0\label{unfcov}
\ee
where
\be
\nabla \equiv d + \half \o^{ab} M_{ab}
\ee
is the Lorentz covariant derivative.

Similarly, it can be shown that taking $\L = \exp (\l^a (x) P_a)$ encodes covariance under local diffeomorphisms, albeit mixed with local Lorentz tranformations (see  \cite{Witten:1988hc} for details).

\subsection{Orthonormal Basis  of Solutions}
The equations of motion (\ref{adsD}) imply that $C$ is a covariantly constant section, and therefore the general solution can be obtained by picking an arbitrary value $C_0 \in R$ to be the value of $C$ in the origin $x=0$ (which we take to correspond  to
the identity,  $G(0) \equiv 1$) and parallel transporting it:
\be
C(x) = G(x)^{-1} C_0\, . \label{gensoladsD}
\ee
Since  the representation $R$ is unitary, the space of solutions to  (\ref{adsD}) has the structure of a Hilbert space,  where the  inner product
is defined as
\be
\left( C  , C'  \right) \equiv \left( C (x) , C' (x) \right)_R = \left( C_0 , C'_0 \right)_R   \label{innprod} \, .
\ee
 Here, $( \cdot , \cdot )_R$ is the  inner product on $R$. The result is independent of $x$ due to (\ref{gensoladsD}) and unitarity.

 If $\{ e_p \}_p$ forms an orthonormal basis of $R$, a complete orthonormal basis of solutions is given by $\{ C_p \}_p$ with
\be
C_p (x) = G^{-1}(x) e_p. \label{basissols}
\ee

\subsection{Global AdS$_{d+1}$ Symmetry}
For a fixed background $\cala$, i.e. a specific choice for the AdS$_{d+1}$ vielbein and spin connection,  the global  symmetries of eq. (\ref{adsD}) are the subset of transformations (\ref{formalgauge1}, \ref{formalgauge2}) which leave  $\cala$ invariant.
From (\ref{gensoladsD}), it is easy to see that  these are generated by infinitesimal gauge parameters of the form
$\l_{AB} =  G^{-1} M_{AB} G$,
which  obviously generate the anti-de-Sitter algebra $so(2,d)$. Their action on the field $C$ is
\be
\d_{AB} C = G^{-1} M_{AB} G \, C \, .\label{symmAdS}
\ee
The basis of solutions (\ref{basissols}) transforms precisely as the representation $R$ of the symmetry algebra:
\be
\d_{AB} C_p = (M_{AB})_p ^{\ q} C_q\, ,
\ee
where the indices $p,q$ refer to components in the representation $R$.
It is therefore clear that  (\ref{adsD}) provides a field theory realization for the particle representation $R$.

Using the equation of motion (\ref{adsD}) and the identity (\ref{Liedef}), we can reexpress the right-hand side of
(\ref{symmAdS}) as the action of the scalar Lie derivative  plus an `internal' part determined by the
Lorentz compensator given in (\ref{KVcomp}):
\be
\d_{AB} C =  l_{AB}^a \pa_a C- \half W_{AB}^{ab} M_{ab} C =  l_{AB}^\m \nabla_\m C+ \half (G^{-1} M_{AB} G)^{ab}M_{ab} C , \label{symmAdSKV}
\ee
where in the second equality we have used (\ref{KVcomp}).
 We note that the second order Casimir differential operator constructed from $\d_{AB}$ is constant, for example for the symmetric tensor representation $D(\D,s)$ it evaluates to
\be
\half \d_{AB} \d^{AB} C = \half M_{AB} M^{AB} C = \left(\D( \D-d) + s(s+ d-2)\right) C \, .
\ee

We end this section with some comments:
\begin{itemize}
\item The unfolded equations (\ref{adsD})  are consistent for general representations $R$, for example the representation ${\bf s}$ in  $D(\D, {\bf s})$ is not restricted to be a symmetric tensor but can have mixed symmetry.
 Though we will focus on the massive case, where $D(\D, {\bf s})$ is a `long' multiplet, in what follows, the above unfolded description  also applies to the massless or partially massless cases, when $D(\D, {\bf s})$ saturates a unitarity bound  and becomes `short'.
The representation $R$ in (\ref{adsD}) could in principle even be non-unitary, though of course in this case the solutions would not form a  Hilbert space.
\item The unfolded description in this section generalizes in a straightforward manner to Minkowski space (the coset Poincar\'e$_{d+1}/ SO(1,d)$) and de Sitter space (the coset $SO(1,d+1)/ SO(1,d)$).
\item While our unfolded equations carry by construction a representation of the AdS$_{d+1}$ symmetry algebra, and therefore also of the simply connected part of the symmetry group, they are not guaranteed to be invariant under additional discrete symmetries (such as parity in $d=2$, as we shall illustrate below). To construct a system invariant under an additional discrete $\ZZ_2$ symmetry may require considering a doublet of fields $C, \tilde C$ which are exchanged by the discrete symmetry.
\end{itemize}
 \section{Unfolded Massive Higher Spin Equations in AdS$_3$}
  Our proposed unfolded equations (\ref{adsD}) give a simple field theory realization of an arbitrary particle representation of the symmetry group. However, they do so at the cost of introducing an infinite number of fields: since unitary representations are infinite dimensional, the field $C$ has an infinite number of components.
Most of these components
 are expected to be in some sense auxiliary, and it will be the goal of this section to understand how to extract the physical components of $C$, focusing on the  case of AdS$_3$ and on massive
 particle representations case for simplicity.  In this case, we will find the explicit linear combinations of components of our master field $C$ which satisfy the topologically massive equations (\ref{topmasseq}). This provides a version of the `central on mass-shell' theorem for our unfolded equations.  In deriving this result, we will also find a map from our master field $C$ to the field obeying the  unfolded equations of \cite{Boulanger:2014vya}.

 \subsection{$sl(2,\RR) \oplus \overline{sl(2,\RR)}$ Basis}
 Let us first specialize the general equations of the previous section to  the case of AdS$_3$.
 The three-dimensional case is somewhat special in that symmetry algebra is not semisimple, $so(2,2)\simeq sl(2,\RR) \oplus \overline{sl(2,\RR)}$, and it will be convenient to work in a basis $L_m, \bar L_m,\ m = -1,0,1$
 adapted to this decomposition. The commutation relations are
 \bea
\, [L_m,L_n] &=&   \e_{mnp}\h^{pq}  L_q = (m-n)L_{m+n} \,,\\
\,  [ \bar L_m, \bar L_n ] &=&   \e_{mnp}\h^{pq} \bar L_q = (m-n)\bar L_{m+n} \,,\\
\, [L_m,\bar L_n] &=&0 \,.
 \eea
 Here, the $\e$-tensor is defined to have $\e_{-1\, 0\,1}=2$  and $\h^{mn}$ is the inverse of
 \be
 \h_{mn} = \left( \begin{array}{ccc} 0&0& -2\\0&1&0\\-2 &0&0 \end{array}\right).
 \ee
The latter is proportional to the Cartan-Killing form which we normalize as
 \be
 K(L_m, L_n) =K(\bar L_m, \bar L_n) = \half \h_{mn}, \qquad K( L_m, \bar L_n)=0.
 \ee
 The generators can be combined into $AdS$-translation generators $P_m$ and Lorentz generators $M_m$, which generate the diagonal $sl(2)$ subalgebra, as follows
 \be
 P_m = L_m - \bar L_m, \qquad M_m = L_m + \bar L_m.
 \ee
 In terms of the original $so(2,2)$ generators $M_{AB},\ A,B = 0',0,1,2$ introduced in (\ref{adsDalg}), these are given by
 \begin{align}
 P_0 &= M_{0'0} \,, & M_0 &= M_{12} \,,\\
 P_{\pm 1} &= M_{0'1}\pm i M_{0'2} \,, & M_{\pm 1} &= M_{02} \mp i M_{01} \,.
 \end{align}
 We note that in unitary representations of $sl(2,\RR)$, the generators must satisfy $L_0^\dagger = L_0$, $L_{\pm 1}^\dagger = L_{\mp 1}$, and similarly for the barred generators. The
  generators of the maximal compact subalgebra $so(2) \oplus so(2)$ are the energy operator $P_0= L_0 - \bar L_0$, which  generates global time translations, and the helicity operator  $M_0= L_0 + \bar L_0$  which generates spatial  $U(1)$ rotations.

  The AdS$_3$ connection splits into $sl(2,\RR)$ and $ \overline{sl(2,\RR)}$ parts:
  \be
  \cala = e^m P_m + \o^m M_m = A^m L_m + \bar A^m \bar L_m,
  \ee
  where
  \be A^m = \o^m + e^m , \qquad \bar A = \o^m - e^m . \label{AAbar} \ee
 Noting that the coset element $G$ splits as
 \be
 G (x) = g (x) \bar g (x),\qquad g \in SL(2,\RR),\ \bar g \in \overline{SL(2,\RR)},
 \ee
 we can work out the equations (\ref{KVcomp}) for the Killing vectors and Lorentz compensator to find
 \begin{align}
 l_m^\m &= - \half (g^{-1} L_m g)^n e_n^\m, & \bar  l_m^\m &=  \half (\bar g^{-1} \bar{L}_m \bar g)^n e_n^\m \label{KVs} \\
 W_m^{\ n} &=  - l_m^\m\bar A_\m^n, &  \bar W_m^{\ n} &=  - \bar l_m^\m A_\m^n. \label{Lorcomp}
\end{align}
 From (\ref{KVs}),  we can derive the following useful identities involving  the Killing vectors:
 \begin{align}
 \h^{mn}l_m^\m l_n^\n &=  {1\over 4} g^{\m\n}, &  \h^{mn}\bar l_m^\m \bar l_n^\n &={1\over 4} g^{\m\n},\label{KVsq}\\
\nabla_{[n} (l_m)_{p]} &=  l_m^\m e_\m ^q \e_{qnp}, & \nabla_{[n} (\bar l_m)_{p]} &= - \bar l_m^\m e_\m ^q \e_{qnp}
 .\label{nablal}
 \end{align}
To derive the identities in the second line we have used the flatness of $\cala$.

 It is a simple exercise to find explicit expressions of the above quantities in the Poincar\'e coordinate system. We take the group elements $g, \ \bar g$  to be
  \be
 g = e^{x_+ L_1} e^{\r L_0},\qquad \bar g = e^{x_- \bar L_{-1}} e^{-\r \bar L_0}.
 \ee
 This leads to
 \begin{align}
 A &=  L_0 d\r +  e^\r L_1 dx_+, & \bar A &=   - \bar L_0 d\r +  e^\r \bar L_{-1} dx_-, \\
 e &=  P_0 d\r + \half  e^\r P_1 dx_+ - \half e^\r  P_{-1} dx_-, & \o &= \half e^\r M_1 dx_+ + \half e^\r  M_{-1} dx_-.
 \end{align}
Computing the metric one indeed finds the AdS$_3$ metric in Poincar\'e coordinates:
 \be
 ds^2 = K(e , e )=  d\r^2 + e^{2 \r} dx_+ dx_-.
 \ee
For the Killing vectors one finds, from (\ref{KVs}),
\begin{align}
 l_{-1} &= e^{-2\r}\pa_- + x_+\pa_\r - x_+^2\pa_+, &\bar l_{-1} &= -\pa_- \,,\\
  l_{0} &=- \half \pa_\r + x_+ \pa_+, &\bar l_{0} &= - x_- \pa_- + \half \pa_\r \,,\\
   l_{1} &= - \pa_+, &\bar l_{1} &=- x_-^2 \pa_- + x_-\pa_\r + e^{-2\r} \pa_+ \,.\label{KVcomps}
 \end{align}

 \subsection{Particle Representations}\label{secreps}
In this section, we will give explicit matrix elements for the unitary representations $D( \D, s)$ in the $sl(2,\RR) \oplus \overline{sl(2,\RR)}$ basis.
We start by reviewing and introducing some  notation for the 
highest- and lowest-weight
representations of the $sl(2,\RR)$ Lie algebra, which are the only ones relevant for our purposes.
 We refer to \cite{Balasubramanian:1998sn}, \cite{Kitaev:2017hnr} for reviews
of  $sl(2,\RR)$ representation theory.

The lowest weight infinite-dimensional  representations  $\cald_+ (h)$, for $2h \neq  \ZZ^-$, are  built on a lowest weight or primary  state   $|0\rangle_h$ satisfying
 \be
 L_1 |0\rangle_h =0, \qquad  L_0 |0\rangle_h= h |0\rangle_h.
 \ee
  If the primary state is normalized, $_h\langle 0|0 \rangle_h =1$, the normalized states in the representation are labelled as $| m\rangle_h, m \in \NN$ and given by
 \be
 | m\rangle_h = \left(m! (2h) (2h +1) \ldots (2h + m-1)\right)^{-\half} (L_{-1})^m |0\rangle_h.\label{Fockbasis}
 \ee
  The generators are represented in this basis as
 \bea
 L_{-1} | m\rangle_h &=&\left((m+1) (2h + m)\right)^{\half}|m+1\rangle_h,\\
 L_0 | m\rangle_h &=& (h+m)| m\rangle_h,\\
 L_1 | m\rangle_h &=& \left(m (2h + m-1)\right)^{\half}| m-1\rangle_h.
 \eea
 For later convenience, we note that the generators can be written in ket-bra notation as
 \bea
 L_{-1} &=& \sum_{m \in \NN} \left((m+1) (2h + m)\right)^{\half}|m+1\rangle_h \,_h\langle m|\\
 L_0 &=& \sum_{m \in \NN} (h+m)| m\rangle_h \,_h\langle m|\\
 L_1 &=& \sum_{m \in \NN} \left((m+1) (2h + m)\right)^{\half}| m\rangle_h \,_h\langle m+1|\, . \label{sl2ketbra}
 \eea
 Using these expressions one checks that, on $\cald_+ (h)$, the quadratic Casimir \be \calc_2 \equiv  \h^{mn} L_m L_n  = L_0^2 - \half (L_1 L_{-1}+ L_{-1}L_1) \ee takes the value
 \be
 \calc_2 = h(h-1)\, . \label{Caslw}
 \ee
 The representations $\cald_+ (h)$  are unitary  for  $h > 0$.

We will also consider the conjugate representations $\cald_-(h)$ for $2h \neq  \ZZ^-$, whose weights are sign-reversed compared to those of $\cald_+ (h)$. These are infinite-dimensional highest weight representations   built on
 a highest weight or anti-primary  state with $L_0$-eigenvalue $-h$, and are unitary for $h>0$. The quadratic Casimir takes again the value (\ref{Caslw}).   The states in these representations can be conveniently denoted as kets $_h\langle  m|$, on which
the $sl(2,\RR)$ generators act from the right with an extra minus sign  to get the right commutation relations. In particular, $_h\langle 0|$ is indeed an anti-primary state satisfying
\be
\,_h\langle 0 | (- L_{-1}) =0, \qquad _h\langle 0 | (- L_{0}) = (-h)  _h\langle 0 |.
\ee

Let us also comment on the cases which were excluded above, built on a lowest  weight state with negative   half-integer weight or a highest weight state with positive half-integer weight.
In this case we obtain a finite-dimensional irreducible representation\footnote{
Note that in  our representations $\cald_+ (h)$ and  $\cald_- (h)$, the generators are manifestly unitarily represented, i.e. $L_m^\dagger = L_{-m}$.  This has the advantage that, when taking the limit where $h$ becomes a negative half-integer,
no null states appear.
This fact will simplify some parts of the subsequent analysis, in particular the results derived in  Appendix \ref{appLorentz}.}  of dimension $2 |h|+1$, which contains both a highest weight $|h|$ and
a lowest  weight $-|h|$ state. The quadratic Casimir takes the value $\calc_2 = |h|(|h|+1)$.
These representations, with the exception of the singlet $h=0$, are non-unitary, and will be denoted by $\cald (|h|)$. They are analytic continuations of the unitary finite-dimensional spin $|h|$ representations of $su(2)$ and we will therefore also refer to them as `spin $|h|$'.

 We are now ready to work out the particle representations of AdS$_3$ in the  $sl(2,\RR) \oplus \overline{sl(2,\RR)}$ basis. Recall
from the previous section that particle representations of $so(2,2)$ are labelled as $D(\D, \h)$ where $\D$ is the energy (eigenvalue of $P_0 = L_0 - \bar L_0$) and $\h$ the helicity
(eigenvalue of $M_0= L_0 +\bar L_0$) of the lowest energy state in the multiplet. From the above considerations, we see that in the $sl(2,\RR) \oplus \overline{sl(2,\RR)}$ basis these
are identified\footnote{A different convention, which often appears in the literature, is related to ours by the redefinition
	$\bar L_m \to - \bar L_{-m}$, which preserves the algebra. In this convention, the particle representations are of the (primary, primary) type $ \left( \cald_+ (h),\cald_+ (\bar h) \right)$, though  $sl(2,\RR) \oplus \overline{sl(2,\RR)}$ is embedded differently into $so(2,2)$, i.e.  $P_m = L_m + \bar L_{-m}, M_m =  L_m - \bar L_{-m}$.}
	as
\be
D(\D, \h) = \left( \cald_+ (h),\cald_- (\bar h) \right)
\ee
where
\be
\D = h + \bar h, \qquad \h = h - \bar h.\label{deltaeta}
\ee
The case where either $h$ or $\bar h$ vanishes describes a short multiplet and corresponds to a massless higher spin particle. We leave the more challenging problem of relating our description of the massless case to the standard Fronsdal equations  for future work, and focus here instead on the case where both $h, \bar h >0$, which corresponds to massive higher spin fields.
The periodicity of the global angular coordinate   furthermore restricts the  helicity $\h$ to be integer (for bosons) or half-integer (for fermions), i.e.
 \be
 h - \bar h \in \ZZ/2.
 \ee
Adopting the
notation where the vectors in  $\cald_-(\bar h)$ are  bra states as discussed above, orthonormal basis states of $ \left( \cald_+ (h),\cald_- (\bar h) \right)$ can be represented as ket-bra states of the form
\be
|m \rangle_{\bar h} \,_{\vphantom{\bar{h}}h}\langle  n |, \qquad m,n \in \NN .\label{ketbras}
\ee
These are orthogonal with respect to the  inner product
\be
 \left( \psi, \psi'  \right) = \tr\, \psi^\dagger  \psi'
 \ee
 where the trace is taken in the $\cald^-_{\bar h}$ Hilbert space.
Note that the states in particle representations can be interpreted as linear maps (or intertwiners) of
$sl(2,\RR )$ representation spaces
\be
 \cald^+_{\bar h} \to \cald^+_{ h}.
\ee

\subsection{Topologically Massive Equations}\label{sectopmass}
Before studying our unfolded equations in more detail, it will be useful to recall a peculiarity of massive higher spin equations in $AdS_3$ which was anticipated in Footnote \ref{fnads3}. In spacetime dimension three, the subgroup of  spatial rotations reduces to $SO(2)$, and the corresponding quantum number is the  helicity $\h = h-\bar h$ in (\ref{deltaeta}). Since parity changes the sign of $\h$, the particle representation $\left( \cald_+ (h),\cald_- (\bar h) \right)$, while furnishing a representation of the component of $SO(2,2)$ connected to identity, is therefore not invariant under parity.  The  Fierz-Pauli equations (\ref{FP}) in AdS$_3$, which
 don't depend on the sign of $\h$ and  are parity-invariant,
actually  describe the direct sum 
\be
\left( \cald_+ (h),\cald_- (\bar h) \right) \oplus \left( \cald_+ (\bar h),\cald_- ( h) \right)
.\ee
The free equations which instead describe only a single helicity $\left( \cald_+ (h),\cald_- (\bar h) \right)$ (and are necessarily parity non-invariant for $\h \neq 0$) are  generalizations of the topologically massive equations for spin one and two \cite{Deser:1981wh}.  It will facilitate our discussion in Section \ref{seclink} below to rederive these equations here from a purely group-theoretic point of view.

We start from a field transforming in the spin-$s$ representation of the Lorentz group, where $s = |\h| = |h-\bar h|$. We can describe this field as
a completely symmetric multi-spinor $\f^{(s)}_{\a_1 \ldots \a_{2 s}}$. The Killing vectors of AdS$_3$ act on it through a generalization of the standard Lie derivative, the so-called Lie-Lorentz derivative (see \cite{Ortin:2002qb} and Appendix \ref{appconvs})
\bea
\call_{l_m} \f^{(s)}_{\a_1 \ldots \a_{2 s}} &=& l_m^\m \left( \nabla_\m \f^{(s)}_{\a_1 \ldots \a_{2 s}} -  s\, e_\m^n (\g_n)_{(\a_1}^{\ \ \ \b} \f^{(s)}_{|\b | \a_2 \ldots \a_{2s})}\right) \,,\nonu 
\call_{\bar l_m} \f^{(s)}_{\a_1 \ldots \a_{2 s}} &=& \bar l_m^\m \left( \nabla_\m \f^{(s)}_{\a_1 \ldots \a_{2 s}} +  s\, e_\m^n (\g_n)_{(\a_1}^{\ \ \ \b} \f^{(s)}_{|\b | \a_2 \ldots \a_{2s})} 
\right) \,.\label{spinorLie}
\eea
In the $\left( \cald_+ (h),\cald_- (\bar h) \right)$ representation, the  $sl(2,\RR)$ and $\overline{sl(2,\RR)}$ Casimirs are equal to
$h(h-1) $ and $\bar h (\bar h-1)$ respectively. Therefore we impose the following field equations:
\be
\left(\h^{mn}  \call_{l_m}\call_{l_n} - h(h-1) \right) \f^{(s)}_{\a_1 \ldots \a_{2 s}} =0, \qquad
\left(\h^{mn}  \call_{\bar l_m}\call_{\bar l_n} - \bar h( \bar h-1) \right) \f^{(s)}_{\a_1 \ldots \a_{2 s}} =0 \, .\label{Casconstr}
\ee
Using the identities  (\ref{KVsq}) for the Killing vectors $l_m^\m, \bar l^\m_m$, these can be rewritten as
\bea
\left(\nabla^\m \nabla_\m - M^2\right)\f^{(s)}_{\a_1 \ldots \a_{2 s}}&=&0, \label{Caseqs1} \\
\nabla_{(\a_1}^{\ \ \ \b}  \f^{(s)}_{|\b|\a_2 \ldots \a_{2 s})}  + \m  \f^{(s)}_{\a_1 \ldots \a_{2 s}}&=&0 \, ,\label{Caseqs}
\eea
where
\be M^2 = \D(\D-2) -s, \qquad \m = {\rm sgn} \h (\D-1).\label{mudef}
\ee
We note that for integer spin the first equation (\ref{Caseqs1}) is the first equation in the   Fierz-Pauli system  (\ref{FP}) in the spinor basis.

For the spin-0 case, the second equation  (\ref{Caseqs}) is actually absent. For $s\neq 0$, the (\ref{Caseqs1},\ref{Caseqs}) equations can be significantly  simplified as follows. It is convenient to introduce an operator $\cald$ which acts on a general multispinor $\t_{\a_1 \ldots \a_{2 s}}$ as
\be
(\cald  \t)_{\a_1 \ldots \a_{2 s}} \equiv \nabla_{\a_1}^{\ \b}  \t_{\b \ldots \a_{2 s}}.
\ee
We should note that, when acting with $\cald$ on a symmetric multispinor the result is in general no longer symmetric, although the square $\cald^2$ does map symmetric tensors into each other. Indeed, one can show using
(\ref{nablacomm}) that
\be
(\cald^2 \f^{(s)})_{\a_1 \ldots \a_{2 s}} = (\Box + s + 1) \f_{\a_1 \ldots \a_{2 s}}. \label{dsqeq}
\ee
Equation (\ref{Caseqs}) can be rewritten as
\be
2 s \m \f^{(s)}_{\a_1 \ldots \a_{2 s}} = (\cald \f^{(s)})_{\a_1 \a_2 \ldots \a_{2 s}}+ (\cald \f^{(s)})_{\a_2 \a_1 \ldots \a_{2 s}} + \ldots + (\cald \f^{(s)})_{\a_{2s} \a_2 \ldots \a_{1}}.
\ee
Acting with $\cald$ on both sides of this equation, the right-hand side is symmetric due to (\ref{dsqeq}), which allows us to derive the integrability condition
\be
\nabla_{[ \a_1} ^{\ \ \ \b} \f^{(s)}_{|\b | \a_2]  \ldots \a_{2 s}} =0.
\ee
This means that the symmetrization in equation (\ref{Caseqs}) can be dropped and we can replace it with
\be
(\cald \f^{(s)})_{\a_1 \ldots \a_{2 s}}   + \m  \f^{(s)}_{\a_1 \ldots \a_{2 s}}=0 .\label{topmasseq}
\ee
These equations replace the full system (\ref{Caseqs1}, \ref{Caseqs}) for $s\neq 0$, since they also imply the Klein-Gordon equation (\ref{Caseqs1}): using  (\ref{dsqeq}) we can write
\be
(\nabla^\m \nabla_\m - M^2) \f^{(s)}_{\a_1 \ldots \a_{2 s}} = ( \cald - \m ) (  \cald + \m ) \f^{(s)}_{\a_1 \ldots \a_{2 s}}.\label{lapldecomp}
\ee
Furthermore, by contracting two indices
in (\ref{topmasseq}), we see that they imply the divergence-free  condition
\be
\nabla^{\b_1 \b_2}  \f^{(s)}_{\b_1\b_2\a_3 \ldots \a_{2 s}}=0
\ee
which for integer spin is precisely the second Fierz-Pauli constraint in (\ref{FP}). The equations (\ref{topmasseq}) therefore imply the Fierz-Pauli equations (\ref{FP}) (and their
generalization for half-integer spin), while it follows from (\ref{lapldecomp})
that the parity-invariant Fierz-Pauli system describes  a pair of topologically massive fields $ \f^{(s)}, \tilde \f^{(s)}$ which satisfy (\ref{topmasseq}) with opposite signs of $\m$, and are exchanged by parity.
The equations (\ref{topmasseq}) are arbitrary spin generalizations \cite{Tyutin:1997yn},\cite{Bergshoeff:2009tb} of the linearized topologically massive spin-1 and spin-2 equations \cite{Deser:1981wh}, and are sometimes referred to as self-dual equations. It can be shown \cite{Deger:1998nm} that they indeed contain the representation $D (h + \bar h, h- \bar h)$.

 \subsection{Unfolded Massive Equations}
After these preliminaries,  let us describe in more detail our unfolded equations (\ref{adsD}) in AdS$_3$.
Our  unfolded master field $C$ is a zero-form taking values in
 the internal space $\left(\cald_+ (h),\cald_- ( h) \right)$, and can be expanded in components in the
ket-bra basis (\ref{ketbras}) as follows
 \be
 C = \sum_{m,n \in \NN} C_{mn} (x) |m \rangle_{\vphantom{\bar{h}}h} \ _{\bar h}\langle  n | 
 \, .\label{comps}
 \ee
 The inner product (\ref{innprod}) on the space of solutions becomes
 \be
 \left( C, C'  \right) = \tr\, C^\dagger (x) C'(x) \, .\label{innprodads3}
 \ee
In terms of the coefficients (\ref{comps}), the inner product equals $\sum_{mn} \bar C_{mn} (x) C'_{mn}(x)$,
 and it is actually independent of $x$ as argued below (\ref{innprod}).

We recall that in the basis (\ref{comps}), the generators of $sl(2,\RR)$ act on $C$  as the operators $L_m$ in the $h$-primary  representation (see (\ref{sl2ketbra})) from the left, while the generators of $\overline{sl(2,\RR)}$ act as the
operators $- L_m$ in the $\bar h$-primary  representation  from the right. In other words, the AdS translations and Lorentz generators act as anticommutators and commutators respectively
\be
P_m C =L_m C + C L_m, \qquad M_m C = L_m  C  - C L_m \, .
\ee
 The unfolded equations  (\ref{unfcov}) read
  \be
 \nabla C + e^m P_m C =0
 \,, \label{eomcov}
 \ee
 where  the Lorentz covariant derivative acts as
\be
\nabla C =\left( d +\o^m M_m \right) C.
\ee
It is sometime useful to write (\ref{eomcov}) in tangent space indices as
 \be
 (\nabla_m + P_m) C=0 \label{unfads3}.
 \ee
We also note that, in terms of the gauge potentials $A = A^m L_m$ and $\bar A = \bar A^m L_m  $ (see (\ref{AAbar}), the equations take a form similar to Vasiliev's  unfolded  equation for the zero form  \cite{Vasiliev:1992gr}
 \be
 d C + A  C - C \bar A =0 \, ,
 \label{eomads3}
 \ee
 although, as we already stressed in the Introduction and will explain in detail below, their group-theoretic content is rather different.

The equations of motion are invariant under the $sl(2,\RR) \oplus \overline{ sl(2,\RR)}$ symmetries of the AdS background which act on the fields as, using  (\ref{symmAdSKV},\ref{Lorcomp}),
\be
\d_{l_m} C = l_m^\m \left( \nabla_\m  - e_\m^m M_m \right) C, \qquad  \d_{\bar l_m} C =\bar  l_m^\m  \left( \nabla_\m  + e_\m^m M_m \right) C.\label{symmads3}
\ee
We note that, just like the topologically massive equation (\ref{topmasseq}),  our unfolded equation is not parity-invariant. Indeed, the natural action of parity on the background gauge fields $A$ is, in Poincar\'{e} coordinates $(t, x, \r)$,
\be
P: \qquad \begin{array}{ccc}A^m_t (t,x,\r) &\to& \bar A^m_t (t,-x,\r) \\ A^m_x (t,x,\r)& \to& - \bar A^m_x (t,-x,\r)\\
A^m_\r (t,x,\r)& \to& \bar A^m_\r (t,-x,\r)\end{array}
\ee
and similarly for $\bar A$, so that $P^2 =1$.
One can check that this leaves the gravitational action $S_{CS} [A] - S_{CS} [\bar A]$ invariant, where $S_{CS}[A]$ is the Chern-Simons action. There is no natural transformation law on $C$ which makes the equation (\ref{eomads3}) invariant under parity. Instead, we can introduce a second field $\tilde C$, taking values in
$(\cald_+ (\bar h), \cald_- ( h))$, with equation of motion
\be
d \tilde C + \bar A \tilde C -  \tilde C  A =0 \, . \label{eomCt}
\ee
The combined system (\ref{eomads3},\ref{eomCt}) is then invariant under the parity transformation
	\be
	C (t,x,\r) \to \tilde C (t,-x,\r)
	\ee
	and similarly for $\tilde C$. The combined system  can also be shown to be time-reversal invariant.

  To make matters a little more concrete,  we can explicitly  work  out some of the  equations in  Poincar\'e coordinates.
 The equations of motion (\ref{eomads3})  read:
  \be
 \pa_\r C + L_0  C + C L_0   =0 , \qquad
 \pa_+ C  + e^\r L_1 C =0, \qquad
 \pa_- C  - e^\r  C L_{-1} =0 .
 \label{eomads3comps}
 \ee
Using (\ref{gensoladsD}), we can also find the general  solution to (\ref{eomads3}). A basis of solutions $\{ C^{[pq]}\}_{p,q}$ is labelled by two natural numbers $p,q$ and obtained by applying the gauge transformation (\ref{gensoladsD}) on constant  basis vectors $|p \rangle_{\vphantom{\bar{h}}h} \ _{\bar h}\langle  q |$ of $(\cald_+(h), \cald_-(\bar h))$. One finds
 \bea
   C^{[pq]} (x) &=& g^{-1}(x) |p \rangle_{\vphantom{\bar{h}}h} \ _{\bar h}\langle  q |  \bar g(x) \nonu
   &=&  e^{-\r( h + \bar h + p + q)}\sum_{j=0}^p \sum_{k=0}^q \caln^{p,q}_{j,k} e^{\r(j+k)}x_+^j x_-^k  |p -j \rangle_{\vphantom{\bar{h}}h} \ _{\bar h}\langle  q-k | \label{Csols} \, , \label{basisads3}
   \eea
  where
   \be
   \caln^{p,q}_{j,k} = (-1)^j \left( {{p}\choose{j}}{{2h + p-1}\choose{j}}{{q}\choose{k}}{{2 \bar h + q - 1}\choose{k}}\right)^\half\, . \label{eq:normFac}
   \ee
   For later reference, let us stress that the solutions (\ref{basisads3}) only have a finite number of non-vanishing components.
   These solutions are by construction orthonormal with respect to the inner product (\ref{innprodads3}):
   \be
(   C^{[pq]},  C^{[p' q']} ) = \d^{pp'} \d^{q q'}\, .
\ee


\subsection{Projecting on Lorentz Tensors}
In this and  the following subsection, we show how our unfolded  equations (\ref{adsD}) are related to other field theory realizations describing the same massive higher spin particle, namely the topologically massive equations (\ref{topmasseq}) and the alternative unfolded description of \cite{Boulanger:2014vya}. Concretely, we will show
that both the topologically massive fields  and the unfolded fields of  \cite{Boulanger:2014vya}  can be constructed as linear combinations of our components fields $C_{mn} (x)$.
The construction relies on interesting group-theoretic properties  which allow  us to project our field $C$ on an irreducible spin-$s$ Lorentz tensor, yielding
the topologically massive equations, or on the $so(2,2)$ representation which underlies the unfolded formulation of \cite{Boulanger:2014vya}.

To illustrate a crucial difference between our unfolded equations and the standard wave equations, it is instructive to compute  the result of acting with the covariant Laplacian on  solutions of our unfolded equations. Using the fact that, on $\left(\cald_+ (h),\cald_- (\bar h) \right)$, we have the Casimir identity
\be
\h^{mn} P_m P_n  + \h^{mn} M_m M_n = \h^{mn} L_m L_n + \h^{mn} \bar L_m \bar L_n =  2  h(h-1) + 2 \bar h (\bar h-1).
\ee
we compute, using the equation of motion (\ref{eomcov}) for $C$,
\be
\nabla_\m \nabla^\m C =
\h^{mn} P_m P_n C=  \left( 2  h(h-1) + 2 \bar h (\bar h-1)  -   \h^{mn} M_m M_n\right) C.
\label{box1}
\ee
 The last term in the brackets is the Casimir operator of the Lorentz subalgebra, which does not evaluate to a constant since our field $C$ transforms in the reducible representation
 $\cald_+(h) \otimes \cald_- (\bar h)$ under the Lorentz subalgebra;
 therefore no component of $C$ itself satisfies a covariant wave equation, in contrast to the standard unfolded formulation \cite{Vasiliev:1992gr},\cite{Barabanshchikov:1996mc},\cite{Boulanger:2014vya}.

 To make contact with covariant wave equations such as the topologically massive equation (\ref{topmasseq}), we should therefore project our master field $C$ onto a field
 $\f^{(s)}$ transforming in the finite-dimensional representation $\cald (s)$, with $s=|h-\bar h|$.
 In other words, we should construct a covariant linear map or intertwiner  between these two Lorentz representations. The existence of such an  intertwiner is somewhat nontrivial,
 as it maps a unitary infinite dimensional representation to a  nonunitary finite-dimensional one. It can therefore not be constructed from the known \cite{repka} Clebsch-Gordan
 decomposition of $\cald_+(h) \otimes \cald_- (\bar h)$  in terms of unitary representations (which involves members of the continuous series with unbounded energies).

 To construct the desired projections we proceed as follows. In Appendix \ref{appLorentz}, we construct  vectors in the $\left(\cald_+(h), \cald_-(\bar h ) \right)$ state space transforming in finite-dimensional representations under the Lorentz
 subalgebra. We find vectors spanning precisely the spin $s+k, \ k \in \NN$, representations which we denote as $V^{(t)}_a$ with $t\geq s, \ |a| \leq t$.
 For $\bar h \geq h$, they are of the form
\be
  V^{(t)}_a = \sum_{n\in \NN} v_n (t,a)  | n \rangle_{\vphantom{\bar{h}}h} \,_{\bar{h}}\langle n- a- s|, \label{Vsmain}
  \ee
 where the coefficients are given in Appendix \ref{appLorentz}, see (\ref{vcoeffs}).
  They transform under the Lorentz subalgebra as
  \be
  M_m  V^{(t)}_a = ( m t + a) V^{(t)}_{a-m} \label{spinsV} \, .
  \ee

We should stress at this point that, as shown in Appendix \ref{appLorentz},   the
vectors $ V^{(t)}_a$ are not normalizeable and are therefore not states in $\left(\cald_+(h), \cald_-(\bar h)\right)$ considered as a Hilbert space. However, it is sufficient for our purposes that they have finite overlap with
the fields $C$ which solve the equations of motion (\ref{eomcov}). This
can be seen by inspecting the
 explicit  solutions (\ref{Csols}):  each basis solution for $C$ has only a finite number of nonzero coefficients.
 Therefore it makes sense to consider the spin-$t$ projections of $C$ defined as the overlap
 \be
 \f^{(t)}_a (x) \equiv \left(   V^{(t)}_a, C(x) \right)\, .\label{spintproj}
 \ee
 These indeed transform in a spin-$t$ representation under Lorentz transformations, since from (\ref{spinsV}) we find the intertwining relation
 \be
   \left(   V^{(t)}_a, M_m C(x) \right) = ( - m t + a) \f^{(t)}_{a+m} (x) \equiv R^{(t)}(M_m)_a^{\ \ b} \f^{(t)}_{b} (x)
 \, . \label{spintrepr}
 \ee
One checks that  $R^{(t)}(M_m) _a^{\ \ b} \equiv (- m t + a ) \d^b_{a +m}$  define basis matrices for  the spin-$t$ representation $\cald(t)$ of the Lorentz subalgebra.

From (\ref{spintrepr}), we can derive a number of useful properties.   First of all,  the spin-$t$ projection of the Lorentz-covariant derivative  $\nabla C$ is precisely the covariant derivative of  $ \f^{(t)}_a (x) $:
 \be
 \left(V^{(t)}_a , \nabla C \right) = d \f^{(t)}_a + \o^m R^{(t)}(M_m)_a^{\ b}  \f^{(t)}_b \equiv \nabla  \f^{(t)}_a \, .
 \ee
 Furthermore, using (\ref{symmads3}), we find that  the spin-$t$ projection of an infinitesimal symmetry transformation acting on $C$ gives precisely the Lie-Lorentz derivative (see (\ref{lieder}, \ref{liederb})) with respect to the corresponding Killing vector:
  \bea
 \left(V^{(t)}_a , \d_{l_m} C \right) &=& l_m^{\m}\left( \nabla_\m \f^{(t)}_a
 - e_\m^p R^{(t)}(M_p)_a^{\ b}  \f^{(t)}_b  \right)\equiv \call_{l_m} \f^{(t)}_a\\
  \left(V^{(t)}_a , \d_{\bar l_m} C \right) &=& \bar l_m^{\m}\left( \nabla_\m \f^{(t)}_a
 + e_\m^p R^{(t)}(M_p)_a^{\ b}  \f^{(t)}_b  \right)\equiv \call_{\bar l_m} \f^{(t)}_a \, .\label{symmKV}
 \eea

The full set of states $V^{(t)}_a$ for $t \geq s, |a|\leq t$ constructed above have the remarkable property that they form an irreducible representation of the full AdS$_3$ symmetry. To show this one needs to check that they transform among themselves under AdS translations. As shown in Appendix \ref{appLorentz},
  this is indeed the case, with
   \be
  P_m  V^{(t)}_a =  2  V^{(t+1)}_{a-m} -{\m s \over t (t+1)} (mt +a)  V^{(t)}_{a-m} +{(s^2-t^2)(\m^2-t^2)\over 2t(2t+1)} d_m(t,a)  V^{(t-1)}_{a-m} \, . \label{Pmaction}
  \ee
  where the coefficients $ d_m(t,a)$ are given in  (\ref{ddef}).
This means that the full set of projections  $ \f^{(t)}_a (x)$ for $t \geq s, |a|\leq t$ also form an irreducible multiplet of  $sl(2,\RR) \oplus \overline{ sl(2,\RR)}$, with translations acting as
 \bea
 ( V^{(t)}_a, P_m C )
&=&  2 \f^{(t+1)}_{a +m}
 -{\m s \over t (t+1)} (- mt +a)  \f^{(t)}_{a+m} +{(s^2-t^2)(\m^2-t^2)\over 2t(2 t+1)} d_{-m}(t,a)  \f^{(t-1)}_{a-m}\nonu &\equiv& P_m  \f^{(t)}_a  \, . \label{Pmeq}
 \eea

 The spin $s=0$ case, $h=\bar h$, deserves a further comment, since  the vectors $ V^{(t)}_a$ constructed above then possess extra structure related to higher spin algebras.
 This extra structure arises because in this case, the particle representation $\left(\cald_+ (h),\cald_- ( h) \right)$
 can be viewed
 as a map from $\cald_+ (h)$ to itself,
 \be
 \cald^+_{ h} \to \cald^+_{ h} \, ,
 \ee
 and it makes sense to consider the product or commutator of the $ V^{(t)}_a$.

 The   state of Lorentz spin 0 is simply the identity operator
 \be V^{(0)}_0 = \sum_n |n\rangle_h \,_h \langle n| = {\bf 1 }
 \ee
 so that the projection on the Klein-Gordon field is simply $\f^{(0)} = \tr C$. The spin-1 vectors are simply the $sl(2,\RR)$ operators
 $ V^{(1)}_m = L_{-m}$ given in (\ref{sl2ketbra}).
 The  lowest weight vector of Lorentz spin $t$ is
 \be
 V^{(t)}_{-t} = (L_1)^t
 \ee
 and the other  $ V^{(t)}_{-t}$ are constructed from these using (\ref{Vnormapp}). By construction, the  $ V^{(t)}_{a}$ for $t\neq$
 form a $hs[\l ]$ algebra under taking  commutators, with $\l^2 = {4 h (h-1)+1}$, while under operator multiplication
 we expect recover the `lone-star' product \cite{Pope:1989sr}. This structure plays a key role in the standard unfolded
 description  of the free massive spin-0 field
  \cite{Vasiliev:1992gr},\cite{Barabanshchikov:1996mc}\footnote{Those works make use of an oscillator realization, which describes the direct sum of two irreducible representations of the symmetry algebra, see  \cite{Raeymaekers:2016mmm} for details, and therefore correspond to a pair of unfolded equations in our approach.  For example, the case $\l = \half$ can be described by a single harmonic oscillator, and gives rise to the direct sum $\left(\cald_+ ({1\over 4}),\cald_- ({1\over 4}) \right) \oplus
  	\left(\cald_+ ({3\over 4}),\cald_- ({3\over 4}) \right)$ . We have checked that the oscillator realization of the $V^{(t)}_s$ is indeed
  	a special case of  our expressions  (\ref{vcoeffs}) for general $h, \bar h $.}.
 Note that, by extending $\cala$ to be an arbitrary flat gauge potential with values in $hs[\l ]$,  the unfolded equations consistently   describe the massive spin-0 field in a background of massless higher spin fields. This can be used to efficiently compute  holographic three-point functions of the scalar-scalar-current type  \cite{Ammon:2011ua}.
 
To recapitulate,  we constructed  through eqs. (\ref{Vsmain}) and (\ref{spintrepr})  an intertwiner between particle representations and tensor representations of the Lorentz algebra for the case of massive higher spin particles in AdS$_3$. We would like to point out that for the case
of  massless  representations in AdS$_D$ with $D \geq 4$,  a similar intertwiner was constructed  in \cite{Iazeolla:2008ix}.

\subsection{Recovering the Topologically Massive Equations}\label{seclink}

With these results in hand, it is now straightforward to show that our unfolded equations imply the topologically massive equations (\ref{topmasseq}). We start by converting
the index $a$ of the  fields $\f^{(s)}_a$ into a rank-$2s$ symmetric spinor index. In our conventions,  this amounts to a simple relabeling of indices, since  our spin-$t$ representation matrices
(\ref{spintrepr}) are precisely the $2t$-th symmetric tensor product of our  spin-$\half$ matrices. Concretely, we define the fields $\f^{(t)}_{\a_1 \ldots \a_{2t}}$, with $\a_j \in  \{ -,+ \}$ as
\be
\f^{(t)}_a \leftrightarrow \f^{(t)}_{\a_1 \ldots \a_{2 t}}
\ee
where the indices are related as
\be
a = \half \sum_{i=1}^{2 t} \a_i, \qquad  \a_1 \ldots \a_{2t} =   \underbrace{+ \dots +}_{\hbox{\footnotesize{$t+a$}}} \underbrace{- \dots -}_{\hbox{\footnotesize{$t-a$}}}
\ee
and we should keep in mind that $\f^{(t)}_{\a_1 \ldots \a_{2t}}$ is defined to be totally symmetric.

Under this relabelling, the symmetry under  Lie-Lorentz derivatives (\ref{symmKV})
gets converted to their equivalent spinorial expressions (\ref{spinorLie}). By construction, the field  $\f^{(s)}_{\a_1 \ldots \a_{2s}}$ satisfies the Casimir relations (\ref{Casconstr}) and, as shown in
Section \ref{sectopmass}, it follows that they also satisfy the topologically massive equations (\ref{topmasseq}).

\subsection{Mapping to the Unfolded System of \cite{Boulanger:2014vya}}\label{seclink2}
We can also show  that the full set of fields $ \f^{(t)}_a (x)$ for $t \geq s, |a|\leq t$ satisfies the unfolded equations of \cite{Boulanger:2014vya}. To this end, we combine (\ref{Pmeq}) with the equation of motion (\ref{eomcov}) to obtain
 \be
\nabla_m  \f^{(t)}_a =  2 \f^{(t+1)}_{a +m}
 -{\m s \over t (t+1)} (- mt +a)  \f^{(t)}_{a+m} +{(s^2-t^2)(\m^2-t^2)\over 2t(2t+1)} d_{-m}(t,a)  \f^{(t-1)}_{a+m}  \, . \label{nablaeq}
 \ee
We  convert this  to spinor form using the identities
\bea
\f^{(t+1)}_{a + m} &=& \half (\g_m)^{\b\g} \f^{(t+1)}_{\b\g \a_1 \ldots \a_{2t}} \\
(- mt + a) \f^{(t)}_{a + m} &=& t  (\g_m)_{(\a_1}^{\ \ \b} \f^{(t)}_{|\b |  \a_2 \ldots \a_{2t})} \\
 d_{-m}(t,a)  \f^{(t-1)}_{a+m} &=&  (\g_m)_{(\a_1 \a_2 } \f^{(t-1)}_{\a_3 \ldots \a_{2t})}.
\eea
The relation (\ref{nablaeq}) therefore becomes
\begin{align}
\nabla_m \f^{(t)}_{\a_1 \ldots \a_{2t}} =&   (\g_m)^{\b\g} \f^{(t+1)}_{\b\g \a_1 \ldots \a_{2t}}  -{\m s \over  (t+1)}  (\g_m)_{(\a_1}^{\ \ \b} \f^{(t)}_{|\b |  \a_2 \ldots \a_{2t})} \nonu &  +{(s^2-t^2)(\m^2-t^2)\over 2t(2t+1)}  (\g_m)_{(\a_1 \a_2 } \f^{(t-1)}_{\a_3 \ldots \a_{2t})}.\label{Bouleqs}
\end{align}
This is, up to convention-dependent normalization factors, precisely the unfolded system of Boulanger et. al. in spinor variables, see eq. (2.28) in \cite{Boulanger:2014vya}. As  was shown there, we can also use (\ref{Bouleqs})  to give an alternative and more direct derivation
of the topologically massive equations (\ref{topmasseq}).
Taking (\ref{Bouleqs})  for $t=s$ and converting the index  $m$ into a pair of spinor indices using (\ref{gammaid}) we obtain
\be
\nabla_{\a\b}  \f^{(s)}_{\a_1 \ldots \a_{2s}} = -2 \f^{(s+1)}_{\a\b \a_1 \ldots \a_{2s}}
+ {\m s\over s+1} \left(\e_{\a (\a_1}\f^{(s)}_{|\b |\a_2 \ldots \a_{2s})} +
\e_{\b (\a_1}\f^{(s)}_{|\a |\a_2 \ldots \a_{2s})} \right) \, .\label{coveqspinor}
\ee
Upon contracting the $\b$ and $\a_1$ indices in this expression, we obtain the
 linearized topologically
massive higher spin equations (\ref{topmasseq}).

The fully symmetrized part of equation (\ref{coveqspinor}) shows that the spin-$s+1$ field $\f^{(s+1)}_{\a_1 \ldots \a_{2s+2}}$ can be obtained by acting with covariant derivatives on the spin-$s$ field $\f^{(s)}_{\a_1 \ldots \a_{2s}}$. By a similar argument holds the same statement holds for components $\f^{(t)}_{\a_1 \ldots \a_{2t}}$ with $t>s$: one can convert the spacetime index $m$ in \eqref{Bouleqs} to spinorial indices. The last two terms in this equation will then be proportional to at least one epsilon tensor and therefore drop out upon considering the fully symmetric component of the equation. All the information is therefore
contained in the spin-$s$ field, while the spin $t>s$ fields are auxiliary.

\subsection{Explicit Mode Solutions}
Using the projections defined above, we can also give a purely algebraic construction of the mode solutions
of the topologically massive gravity equations (\ref{topmasseq}).  In Poincar\'{e} coordinates, we start from   the basis of  solutions $\{ C^{[pq]} \}_{p,q\in \NN}$ of (\ref{basisads3}) and work out the  projection on a   spin-$s$ tensor (assuming $\bar h \geq h$) using (\ref{spintproj}, \ref{Vsmain}) to find

\bea
\f^{(s) [pq]}_a &\equiv & \left(V^{(s)}_a ,  C^{[pq]} \right)\nonu
&=& e^{(a + s - \D)\r} \sum_{n=0}^\infty v_n (s,a) \caln^{pq}_{p-n,q+a + s-n} e^{- 2n \r}x_+^{p-n} x_-^{q+a+s-n}\label{modesol} \,,
\eea
where the normalization factor $\mathcal{N}$ was defined in \eqref{eq:normFac} and the explicit expression for $v_n$ is given in Appendix~B, see \eqref{vcoeffs}.

 These solutions form a basis of the topologically massive
equations transforming as $\left( \cald_+ (h),\cald_- (\bar h) \right)$ under the AdS$_3$ symmetries.  A potential caveat in our  reasoning so far was that the spin-$s$ projections of our solutions might actually vanish; the above expression shows this not to be the case. Indeed, they yield the full multiplet of solutions to the topologically massive equations. For example, for the lowest component $\f^{(s)[pq]}_{-s} = \f^{(s)[pq]}_{--\ldots -}$ (\ref{modesol}) reduces to, up to a normalization factor,
\be
 \f^{(s)[pq]}_{-s} \sim e^{-( h + \bar h ) \r} \, x_+^{p} x_-^{q} \; _2F_1 \left(-p,-q, 2 h, -{e^{-2 \r}\over (x_- x_+)^2}\right).
\ee
 This expression is nonvanishing and finite since the hypergeometric function truncates to a polynomial with a finite number of terms.
One checks that for $h = \bar h$ the physical field satisfies the Klein-Gordon equation with the correct mass term (\ref{FPmass}):
\be
( \nabla^\m \nabla_\m - 4 h(h-1) ) \f^{(0)[pq]}_{0} =0.
\ee

  \section{Outlook}
  In this work, we have proposed a simple unfolded description of particles in an arbitrary representation of the spacetime symmetry.
  It is somewhat nontrivial to  connect this formulation with the standard covariant wave equations, which requires one to construct an intertwiner between this  representation and the appropriate tensor under the Lorentz algebra, which we worked explicitly for massive fields of arbitrary spin in AdS$_3$ (see \cite{Iazeolla:2008ix} for the case of massless
  fields in AdS$_{D \geq 4}$). One could say that in this approach, the problem of unfolding a given relativistic wave equation reduces to the  representation theoretic problem of constructing the appropriate intertwiner.
  
 We end by pointing out some open problems and possible generalizations.
  \begin{itemize}
  \item Our construction of the projection on spin-$t$ tensors, using the non-normalizeable vectors $ V^{(t)}_a $, was somewhat pedestrian and deserves a more rigorous treatment. This could also elucidate whether our unfolded equation is truly equivalent to the alternative unfolded formulation of \cite{Boulanger:2014vya}: though we found a map from our master field to the one in  the formulation of \cite{Boulanger:2014vya}, it is not clear if this map is invertible  (see  Appendix A of \cite{Campoleoni:2013lma}
  for a discussion of this issue in the spin-0 case).
  \item As we show in Appendix \ref{appLorentz}, it is possible to construct highest weight (for $h > \bar h $) or lowest weight (for $h < \bar h $)  vectors   with Lorentz spin lower than $s$ in the representation space $( \cald_+ (h) , \cald_- (\bar h))$.
 These are however not part of an irreducible Lorentz representation, rather they form an indecomposable structure. The meaning of the projection of our unfolded field $C$ on this  indecomposeable structure is unclear to us, though it is somewhat suggestive of a dual formulation involving gauge fields.
  \item Our explicit construction for AdS$_3$ could be generalized in various ways. For example, in higher dimensions one might expect   to be able to construct an intertwiner between the particle representation $D (\D , s)$ and the multiplet that underlies
  the unfolded massive equations of \cite{Ponomarev:2010st}. It would also be interesting to  study, using the results of \cite{Iazeolla:2008ix}, the relation between our formulation for  massless
  particles and the Fronsdal equations and their standard unfolded form \cite{Lopatin:1987hz},   and to study the partially massless case. 
  \item One of our motivations for studying the current unfolded formulation is that it arises naturally  in the AdS$_3$
  theory with  higher spin square gauge symmetry. In a separate publication \cite{inprogr}, we will show that the natural equation describing matter
 coupled to the higher spin square \cite{Raeymaekers:2016mmm} describes  an infinite set of unfolded massive higher spin equations of the type studied in this work.
  \item  Since the  present unfolded  formulation has a   clear group theoretic meaning which involves only the particle representation $D(\D , s)$, it may be hoped that
  it provides a natural framework to describe higher spin interactions. It would be interesting to give a more group-theoretic characterization of the interaction vertices  in Vasiliev theory, especially in their recently developed local form \cite{Vasiliev:2016xui}, in our framework. It may be also be hoped
  that the current setup is the natural one for addressing the open problem of constructing the fully interacting theory with higher spin square gauge symmetry.

  \end{itemize}

 \acknowledgments
  
  We thank  Stefan Fredenhagen, Carlo Iazeolla, Tom\'a\v{s} Proch\'azka, Evgeny Skvortsov and Misha Vasiliev for useful discussions. We are indebted to Mitya Ponomarev and Nicolas Boulanger for carefully reading our manuscript. P.K. would like to thank the Albert Einstein Institute for generous support by which his contribution to the present work became possible.
 The research of J.R.
  was supported by the Grant Agency of the Czech Republic under the grant 17-22899S.

\begin{appendix}
\section{AdS$_3$ Conventions}\label{appconvs}
In this appendix, we spell out some of our conventions for AdS$_3$. The
AdS$_3$ symmetry algebra is
\be
\, [M_m, M_n ]= \e_{mn}^{\ \ \  p} M_p,\qquad
 [M_m, P_n ] = \e_{mn}^{\ \ \  p} P_p,\qquad
 [P_m, P_n ] = \e_{mn}^{\ \ \  p} P_p,\label{algapp}
\ee
where
\be
\h_{mn} = \left( \begin{array}{ccc} 0&0& -2\\0&1&0\\-2 &0&0 \end{array}\right), \qquad \e_{-0+}\equiv 2.
\ee
In terms of the $sl(2,\RR) \oplus \overline{sl(2,\RR)}$ basis we have $M_m = L_m +\bar L_m, \ P_m=  L_m -\bar L_m$.
From the dreibein and spin connection, we can form the
AdS$_3$ connection
\be
\cala = e^m P_m + \o^m M_m
\ee
whose flatness, $d \cala + \cala \wedge \cala =0$, is equivalent to the structure  equations
\be
d e^m + \e_{\ np}^{m} e^n \wedge \o^p = 0, \qquad
d \o^m + \half \e_{\ np}^{m} ( \o^n \wedge \o^p + e^n \wedge e^p ) = 0.\label{curveq}
\ee

 Let $\F_R$ be a field transforming in a representation $R$ under local Lorentz tranformations.
The Lorentz covariant derivative is
\be
\nabla_\m \F_R = (\pa_\m + \o_\m^m R(M_m) ) \F_R.
\ee
where $ R(M_m)$ are the representation matrices. For example, on a tangent vector  the appropriate representation is
\be
 R(M_m) v^n = - \e_{m\ \ p}^{\ \  n} v^p.\label{vectrep}
\ee
The covariant derivative can be extended to  tensors with curved indices in the usual way   using the Christoffel symbols.
 The covariant derivative has the following properties
\bea
\nabla_\m e_\n^m &=&0 \,,\\
\nabla_\m (R( M_m) \F_R) &=& R( M_m) \nabla_\m  \F_R \,,\\
\, [\nabla_m , \nabla_n ] \F_R &=& - \e_{mn}^{\ \ \ p} R(M_p) \F_R \,,\label{nablacomm}
\eea
where the last identity is derived from  (\ref{curveq}).

We often use spinor notation, with indices $\a, \b, \ldots \in \{-, +\}$, which are raised and lowered with $\e_{\a\b}$ and $\e^{\a\b}$, where
\be
\e_{-+ } =\e^{-+} =1.
\ee
We use `northwest-southeast'  conventions:
\be
\e^{\a\b}v_\b=v^\a, \qquad  v^\b \e_{\b\a} = v_\a\, .
\ee
The gamma matrices are denoted as $\g_m, \ m \in \{ -,0 ,+\}$ are given by
\be
(\g_-)_\a^{\ \b} = \left( \begin{array}{cc}  0&0\\2 &0\end{array} \right), \qquad
(\g_0)_\a^{\ \b} = \left( \begin{array}{cc}  -1&0\\ 0 &1\end{array} \right),\qquad
(\g_+)_\a^{\ \b} = \left( \begin{array}{cc}  0&-2\\ 0 &0\end{array} \right)\, ,
\ee
and they satisfy
\be
\g_m \g_n = \h_{mn}+  \e_{mn}^{\ \ \  p} \g_p, \qquad (\g^m)_{\a_1 \a_2}  (\g_m)^{\b_1 \b_2} =-2\d^{(\b_1}_{\a_1}\d^{\b_2)}_{\a_2}.\label{gammaid}
\ee

The spin-$s$  representation of the Lorentz algebra acts on a rank-$2s$ symmetric multi-spinor as
\be
R_{s} (M_m )\f_{\a_1 \ldots \a_{2s}} = s (\g_m)_{(\a_1}^{\ \ \ \b} \f_{|\b | \a_2 \ldots \a_{2s})}.
\ee
In the main text we make use of an  operator $\cald$ defined as
\be
(\cald  \f)_{\a_1 \ldots \a_{2 s}} \equiv \nabla_{\a_1}^{\ \b}  \f_{\b \ldots \a_{2 s}}
\ee
one can show, using (\ref{nablacomm}), that
\be
\cald^2 \f_{\a_1 \ldots \a_{2 s}} = (\Box + s + 1) \f_{\a_1 \ldots \a_{2 s}} \,. \label{dsqeqapp}
\ee
Note that $\cald$  does not map symmetric spinors into symmetric spinors in general, while $\cald^2$ does.

We also make use of the Lorentz-covariant  definition of the Lie derivative with respect to a Killing vector $k$,
  acting on fields in  arbitrary representations of the Lorentz algebra, see  \cite{Ortin:2002qb}.
  This definition extends the standard  definition of the Lie derivative of tensor fields and is called the Lie-Lorentz derivative.
  For a field $\f_a$ transforming in a representation $R$ under the Lorentz algebra, it is defined as
\be
\call_k  \f_a = k^\m \nabla_\m   \f_a + \half \nabla_{[m }k_{n]} \e^{mnp} R (M_p)_a^{\ b}\f_b \,.
\ee
From this definition, using (\ref{vectrep}) and the fact that $k$ is a Killing vector, one shows that
\be
\call_k e_\m^m =0, \qquad \call_k R(M_m)_a^{\ b} =0.
\ee
On AdS$_3$, using the   identities (\ref{nablal}),
the Lie-Lorentz derivative simplifies to
\bea
\call_{l_m} \f_a  &=& l_m^{\m}\left( \nabla_\m \f_a
- e_\m^p R(M_p)_a^{\ b}  \f_b  \right) \,, \label{lieder}\\
\call_{\bar l_m} \f_a   &=& \bar l_m^{\m}\left( \nabla_\m \f_a
+ e_\m^p R(M_p)_a^{\ b}  \f_b  \right) \, .\label{liederb}
\eea
In spinor notation, this leads to (\ref{spinorLie}).

\section{Representations of the Lorentz Subalgebra}\label{appLorentz}

In this appendix, we will construct vectors in the representation space\footnote{ Viewed here as the vector space $Span\{ | m\rangle_{\vphantom{\bar{h}}h} \,_{\bar{h}}\langle n|, m, n \in \NN \}$, in particular  we will not insist on normalizeability with respect to the norm (\ref{innprodads3}) on  $( \cald_+(h) , \cald_-({\bar h}))$.} $( \cald_+(h) , \cald_-({\bar h}))$ with $h, \bar h >0$ which transform in finite-dimensional representations of the Lorentz subalgebra.
We recall that, in the ket-bra notation introduced in Section \ref{secreps}, the Lorentz generators $M_m$ act on this space as $M_m \psi = L_m \psi - \psi L_m$, with the $L_m$ given explicitly in (\ref{sl2ketbra}). We start by constructing all lowest Lorentz  weights
$\l_w$ and highest weights $\n_w$ defined by the properties
\begin{align}
M_0  \l_w &= w  \l_w, & M_1 \l_w &= 0\\
M_0  \n_w &= w  \n_w, & M_{-1} \n_w &= 0.
\end{align}
It follows that if $\l_w$ is a lowest weight $w$ state, then
\be
\l_{w-k} \equiv L_1^k \l_w = \left({P_1 \over 2}\right)^k \l_w, \qquad k \in \NN \, ,\label{P1act}
\ee
if nonvanishing, is another lowest weight  state of weight $w-k$. Similarly if $\n_w$ is a highest weight $w$ state, then
\be
\n_{w+ \tilde k} \equiv L_{-1}^{\tilde k} \n_w = \left({P_{-1} \over 2}\right)^{\tilde k} \n_w, \qquad \tilde k \in \NN  \, ,\label{Pm1act}
\ee
if nonvanishing, is another highest weight  state of weight $w + \tilde k$. Starting from an ansatz describing the most general vector with fixed Lorentz weight under $M_0$, one finds that the full set of  lowest (highest) weight vectors can be obtained in this way, through repeated action  of
of $L_1$ ($L_{-1}$) on a single starting vector.

One finds lowest weight vectors at weights $- s -k$  and highest  weight vectors at weights $- s + \tilde k$, with $k, \tilde k \in \NN$ and $s \equiv |h - \bar h|.$ Assuming for the moment that  $h \leq \bar h$ (we will return to the case $h > \bar h$ below),
these vectors are given by
\begin{align}
\l_{-s-k} &= L_1^k \l_{-s}, & \l_{-s} = \sum_{n \in \NN} \left( {( 2\bar h + n -1)!\over
	(2 h + n -1)!}\right)^\half  | n \rangle_{\vphantom{\bar{h}}h} \,_{\bar{h}}\langle n|,\label{lowestw}\\
\n_{-s+ \tilde k} &= L_{-1}^{\tilde k} \n_{-s}, & \n_{-s} = \sum_{n \in \NN} \left( {( 2 h + n -1)!\over
	(2 \bar h + n -1)!}\right)^\half  | n \rangle_{\vphantom{\bar{h}}h} \,_{\bar{h}}\langle n|\, .
\label{highestw}
\end{align}

\begin{figure}[h]
\centering
\includegraphics[width=200pt]{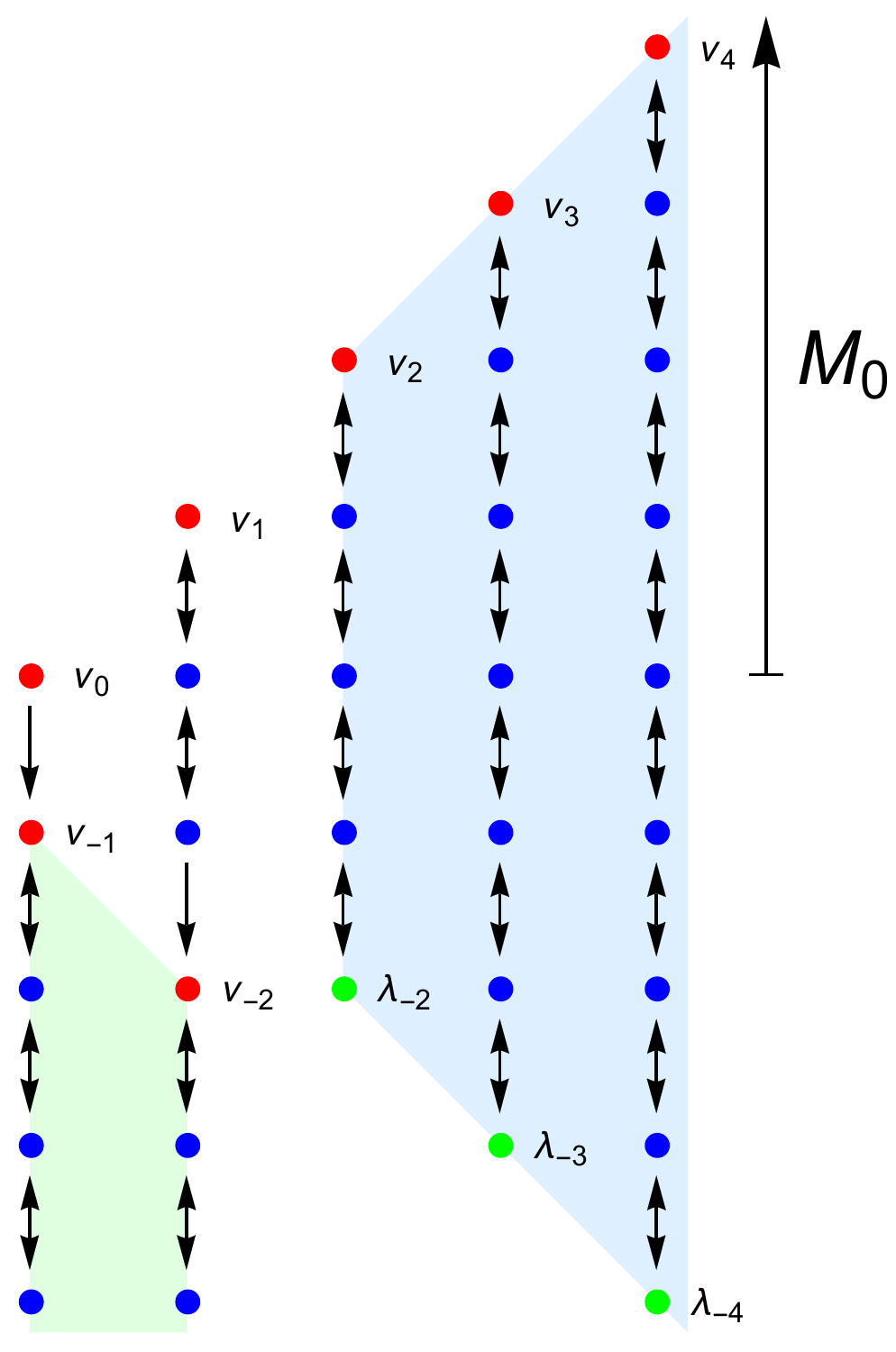}
\caption{Weight diagram showing the highest (in red) and lowest (in green)  weight states under the Lorentz algebra and their descendants (in blue), for the  spin 2 case with $\bar h - h =2$. An arrow pointing up (down) means that the states are linked by the action of $M_{-1}$ ($M_{1}$). We note that the states $\n_{-1}$ and $\n_{-2}$ are null primaries. The states in the blue shaded region are the $V^{(t)}_a$ which form an irreducible representation of the full AdS algebra $sl(2,\RR) \oplus \overline{sl(2,\RR)}$. Only the  states in the green shaded  region are actually normalizeable. }
\label{fig:weights}
\end{figure}

Let us now proceed to arrange these highest and lowest weight vectors and their Lorentz descendants into  multiplets, see
 Figure \ref{fig:weights} for the resulting weight diagram in the case of spin 2.

Setting $ \tilde k = 2 s + k$ in (\ref{highestw}), we find that $\n_{s + k} \sim
M_{-1}^{ 2(s + k)} \l_{-s-k}$ and hence these vectors are the lowest and highest weights of a
$2(s + k)+1$-dimensional  spin-$(s + k)$ representation, denoted as $\cald (s+k)$ in Section \ref{secreps}. We will  denote the basis vectors in the $\cald (t)$ multiplet as $V^{(t)}_a$,
with $a$ running from $-t $ to $t$ and normalized as follows:
\be
 V^{(t)}_a = (-1)^{t+a} {(t-a)! \over (2 t)!} M_{-1}^{t+a} \left( L_1^{t-s} \l_{-s} \right), \qquad t \geq s, |a|\leq t \label{Vnormapp}
 \ee
Our normalization constants are chosen such that $ V^{(t)}_{-t} = \l_{-t},  V^{(t)}_{t} = \n_{t}$ and such that the Lorentz generators act as
\be
M_m  V^{(t)}_a =(mt + a)  V^{(t)}_{a-m}\, .\label{LorVapp}
\ee
  For the explicit component expression of the $ V^{(t)}_a$ one finds
  \be
  V^{(t)}_a = \sum_{n\in \NN} v_n (h, \bar h; t,a)  | n \rangle_{\vphantom{\bar{h}}h} \,_{\bar{h}}\langle n- a- s|, \qquad {\rm for\ } h\leq \bar h\label{Vs}
  \ee
  with (recall $\D\equiv h + \bar h$)
  \begin{align}
  v_n (h, \bar h;t,a)   =&  {(t-a)! \over (2 t)!} \sum_{l=0}^{t+a} (-)^{l} \binom{t+a}{l} \Big( (1-2h-n)_l (-n)_l (1-l+n)_{t-s} (2h-l+n)_{t+s} \nonu &  (2-a-s+n)_{t+a-l-1} (1+\D-a+n)_{t+a-l-1} (1-a-s+n) (\D-a+n) \Big)^{1/2} \,, \label{vcoeffs}
  \end{align}
where $(x)_n=x (x+1) \dots (x+n-1)$ denotes the Pochhammer symbol. To find similar expressions for the case $h > \bar h$, one notes that the Hermitean conjugate $(V^{(t)}_a)^\dagger$ is a state of weight $-a$ in the space  $( \cald_+(\bar h) , \cald_-({ h}))$, leading to
 \be
V^{(t)}_a = \sum_{n\in \NN} v_n (\bar h,  h; t,-a)  | n+ a - s \rangle_{\vphantom{\bar{h}}h} \,_{\bar{h}}\langle n|, \qquad {\rm for\ } h\geq \bar h
\ee

As we saw in (\ref{LorVapp}), the vectors $V^{(t)}_a$ span the representation $\oplus_{t=s}^\infty \cald(t)$ under the Lorentz subalgebra generated by $M_m$. What is more, they also transform among themselves
under the full $sl(2,\RR) \oplus \overline{ sl(2,\RR) }$ symmetry, under which they form a single  irreducible representation. To see
this, we have to work out how the AdS translation generators act on them; one can derive the following relation:
 \be
P_m  V^{(t)}_a =  2  V^{(t+1)}_{a-m} -{\m s \over t (t+1)} (mt +a)  V^{(t)}_{a-m} +{(s^2-t^2)(\m^2-t^2)\over 2t(2 t+1)} d_m(t,a)  V^{(t-1)}_{a-m} \, . \label{Pmactionapp}
\ee
where $\m$ was defined in (\ref{mudef}) and we defined the coefficients
\be
d_\pm (t,a) ={(a \pm t)(a \pm (t-1))\over t (2t-1)}, \qquad d_0 (t,a) =  {a^2-t^2 \over t(2t-1)} \,.\label{ddef}
\ee
The coefficients in (\ref{Pmactionapp}) are completely fixed by the properties (\ref{P1act},\ref{Pm1act}), consistency with the AdS algebra (\ref{algapp}) and the Casimir relation
\be 
\h^{mn} P_m P_n V^{(t)}_a = \left( 2h(h-1) + 2 \bar h (\bar h -1) - t(t+1) \right)  V^{(t)}_a \, .
\ee
We also verified (\ref{Pmactionapp}) using the explicit expressions (\ref{Vs}).

We also mention a further useful relation involving the action of the translation generators $P_m$ on the vectors $V^{(t)}_a$. Recalling that these act as $P_m \psi = L_m \psi + \psi L_m$
one can derive the following identity
\be
M_{-1}^{n+1} P_1 = n(n+1) P_{-1} M_{-1}^{n-1} - 2(n+1) P_0 M_{-1}^n + P_1 M_{-1}^{n+1}  \, .
\ee
Combining this with  (\ref{Vnormapp}), we find a  relation expressing the vectors in the spin-$t+1$ representation in terms of the action of the translation generators on vectors in the spin-$t$ representation:
\be
V^{(t+1)}_a =\sum_{m=-1}^1  c_m (t,a)  P_{m} V^{(t)}_{a+m}  \label{spinraise}
\ee
where
\be c_{-1} =  { (t+a)(t+a+1) \over 2(2t+2)(2 t + 1)}, \qquad c_0 ={  (t+a+1)(t-a+1) \over (2t+2)(2 t + 1)},
\qquad c_1 ={(t-a+1)(t-a) \over 2(2t+2)(2 t + 1)}\, . \label{spinraisecoeff}
\ee

In (\ref{highestw}) we also found, for $\bar h > h$,  a number of highest weight vectors, namely $\n_{-s + \tilde k}$ for $0 \leq \tilde k < 2 s$, for which there is no corresponding lowest weight vector and  which hence do  not fit  in finite-dimensional representations.
  Though these do  not play a  role in the present work (see however the Outlook section), we now briefly comment on the Lorentz representations carried by these states. One checks that
    $\n_{-s + \tilde  k}$ for $0\leq \tilde k < s - \half$ is  actually a Lorentz descendant of $\n_{s-\tilde k-1}$:
  \be
  \n_{-s + \tilde k} \sim M_1^{2(s-\tilde k)-1} \n_{s-\tilde k-1} \qquad  {\rm for\ }  0\leq k < s - \half\, .
  \ee
   The $ \n_{-s + \tilde   k}$  for $0\leq\tilde  k < s - \half$ and their descendants therefore form an invariant subspace whose complement is not invariant; in that case $\n_{-s +\tilde  k}$ and $\n_{s-\tilde k-1} $   belong to an infinite-dimensional reducible but indecomposable representation of the Lorentz subalgebra.

Let us also discuss the normalizablility of the vectors constructed above with respect to the
inner product (\ref{innprodads3}). One finds that the norm of the highest weight vectors for $ \bar h \geq h$  is
\be
| \n_{-s + \tilde  k}|^2 = {k! (2  h +\tilde  k -1)! \over (2\bar h -1)!}   \,_2 F_1 (\tilde k+1 , 2 h +\tilde k, 2 \bar h,1),\qquad {\rm for \ } h\leq \bar h
\ee
Using well-known convergence properties of the hypergeometric function \cite{abramowitz}, we find that
the highest weight states are therefore normalizable for $\tilde  k < s - \half $. In other words, only
the `null' highest weight vectors discussed above are actually normalizeable\footnote{The representations built on these normalizeable highest weight vectors are precisely the ones appearing in the Clebsch-Gordan decomposition of $\cald_+(h) \otimes \cald_-(\bar h)$ in terms of unitary representations, see \cite{repka}.}. In particular, the vectors $ V^{(t)}_a$ comprising the finite-dimensional representations are not normalizable.

\end{appendix}

\end{document}